\documentclass[11pt,aps,pre,preprint,floatfix,superscriptaddress]{revtex4-1}

\usepackage{amsmath}
\usepackage{amssymb}
\usepackage{graphicx}
\usepackage{natbib}
\usepackage[usenames,dvipsnames,svgnames,table]{xcolor}
\usepackage{txfonts}

\def\l{\left}
\def\r{\right}
\newcommand{\f}{\frac}

\newcommand{\deff}{d_\mathrm{eff}}

\newcommand{\qc}{q_{\mathrm{c}}}

\newcommand{\avg}[1]{\ensuremath{\left<#1\right>}}
\newcommand{\dnn}{\langle d_{\mathrm{nn}}\rangle}
\newcommand{\knn}{\langle k_{\mathrm{nn}}\rangle}

\newcommand\given[1][]{\:#1\vert\:}

\date{\today} 

\begin{document}

\title{Robustness of Spatial Micronetworks}

\author{Thomas C.~McAndrew}
\email{Thomas.McAndrew@uvm.edu}
\homepage{www.uvm.edu/~tmcandre/}
\affiliation{Department of Mathematics and Statistics University of Vermont, Burlington, VT, United States}
\affiliation{Vermont Complex Systems Center, University of Vermont, Burlington, VT, United States}

\author{Christopher M.~Danforth}
\affiliation{Department of Mathematics and Statistics University of Vermont, Burlington, VT, United States}
\affiliation{Vermont Complex Systems Center, University of Vermont, Burlington, VT, United States}

\author{James P.~Bagrow}
\affiliation{Department of Mathematics and Statistics University of Vermont, Burlington, VT, United States}
\affiliation{Vermont Complex Systems Center, University of Vermont, Burlington, VT, United States}

\begin{abstract}
Power lines, roadways, pipelines and other physical infrastructure are critical to modern society. These structures may be viewed as spatial networks where geographic distances play a role in the functionality and construction cost of links. 
Traditionally, studies of network robustness have primarily considered the connectedness of large, random networks. Yet for spatial infrastructure physical distances must also play a role in network robustness. 
Understanding the robustness of small spatial networks is particularly important with the increasing interest in microgrids, small-area distributed power grids that are well suited to using renewable energy resources.
We study the random failures of links in small networks where functionality depends on both spatial distance and topological connectedness. By introducing a percolation model where the failure of each link is proportional to its spatial length, we find that, when failures depend on spatial distances, networks are more fragile than expected.
Accounting for spatial effects in both construction and robustness is important for designing efficient microgrids and other network infrastructure.
\end{abstract}

\maketitle


\section{Introduction}

 
 %
 
 
 




The field of complex networks has grown in recent years with applications across many scientific and engineering disciplines~\cite{albert2002statistical,newman2003structure,newman2010networks}.
%
Network science  has generally focused on how  topological characteristics of a network affect its structure or performance~\cite{albert2002statistical,callaway2000network,newman2003structure,strogatz2001exploring,crucitti2004topological,newman2010networks}. 
Unlike purely topological networks, spatial networks~\cite{barthelemy2011spatial} like roadways, pipelines, and the power grid must take physical distance into consideration.
Topology offers indicators of the network state, but ignoring the spatial component may neglect a large part of how the network functions ~\cite{watts2002simple,ball1997epidemics,buldyrev2010catastrophic,albert2004structural}. 
%
For spatial networks in particular, links of different lengths may have different costs affecting their navigability~\cite{kleinberg2000navigation,roberson2006kleinberg,campuzano2008kleinberg,carmi2009asymptotic,PhysRevLett.102.238703,PhysRevLett.104.018701} and construction~\cite{PhysRevLett.89.218701,gastner2006spatial,gastner2006shape,tero2010rules}.
Percolation~\cite{stauffer1991introduction} provides a theoretical framework to study how robust networks are to failure~\cite{callaway2000network,newman1999scaling,bagrow2011robustness,shekhtman2014robustness,schneider2011mitigation}. In traditional bond percolation, each link in the network is independently removed with a  constant probability, and it is asked whether or not the network became disconnected. Theoretical studies of percolation generally assume very large networks that are locally treelike, often requiring millions of nodes before finite-size effects are negligible. Yet many physical networks are far from this size; even large power grids may contain only a few thousand elements.

There is a need to study the robustness of small spatial networks. Microgrids~\cite{lasseter2004microgrid,smallwood2002distributed,hatziargyriou2007microgrids,katiraei2006power} are one example. Microgrids are small-area (30--50 km), standalone power grids that have been proposed as a new model for towns and residential neighborhoods in light of the increased penetration of renewable energy sources. Creating small robust networks that are cost-effective will enable easier introduction of the microgrid philosophy to the residential community. 
Due to their much smaller geographic extent, an entire microgrid can be severely  affected by a single powerful storm, such as a blizzard or hurricane, something that is unlikely to happen to a single, continent-wide power grid.
Thus building on previous work, we consider how robustness will be affected by spatial and financial constraints. The goal is to create model networks that are both cost-effective, small in size, and at the same time to understand how robust these small networks are to failures.
%


The rest of this paper is organized as follows. In Sec.~\ref{sec:infrastructureModel} a previous model of spatial networks is summarized. Section~\ref{sec:robustPhysInf} contains a brief summary of percolation on networks, and applies these predictions to the spatial networks. In Sec.~\ref{sec:modelInfRob} we introduce and study a new model of percolation for spatial networks as an important tool for infrastructure robustness. Section~\ref{sec:discuss} contains a discussion of these results and future work.


\section{Modeling infrastructure networks}
\label{sec:infrastructureModel}

\begin{figure}
{\includegraphics[trim=35 60 30 25,clip=True]{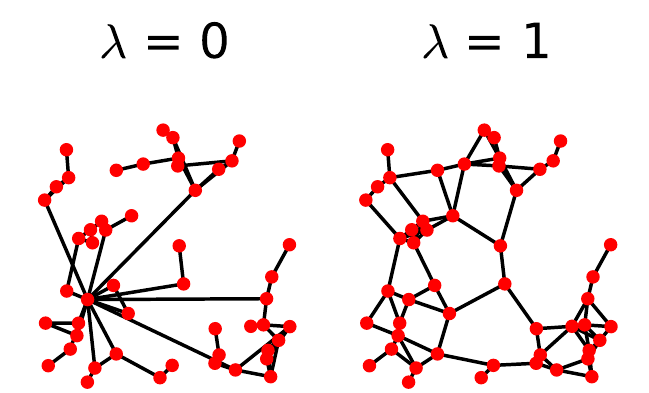}}
\caption{(Color online) Two optimized spatial networks with the same node coordinates illustrate how $\lambda$ influences network topology. The \textbf{non-spatial} case $\lambda = 0$ shows long-range hubs due to the lack of restriction on edge distance; the \textbf{spatial} case $\lambda = 1$ lacks expensive long distance links leading to a more geometric graph. As examples, the non-spatial case may correspond to air travel  where minimizing the number of flights a traveler takes on a journey is more important than minimizing the total distance flown, while the spatial case may represent a road network where the overall travel distance is more important than the number of roads taken to reach a destination.
\label{samecoords}
}
\end{figure}

In this work we consider a spatial network model introduced by Gastner \& Newman~\cite{gastner2006shape,gastner2006spatial}, summarized as follows. A network consists of $|V| = N$ nodes represented as points distributed uniformly at random within the unit square. Links are then placed between these nodes according to associated construction costs and travel distances. The construction cost is the total Euclidean length of all edges in the network, 
   $ \sum_{ (i,j) \in E}{d_{ij}}$,
where $d_{ij}$ is the Euclidean distance between nodes $i$ and $j$ and $E$ is the set of undirected links in the network. This sum represents the capital outlay required to build and maintain the network. When building the network, the construction cost must be under a specified budget.
Meanwhile, the travel distance encapsulates how easy it is on average to navigate the network and serves as an idealized proxy for the functionality of the network. The degree to which spatial distance influences this functionality is tuned by a parameter $\lambda$ via an ``effective'' distance
	\begin{equation*}
		\deff(i,j) = \sqrt{N}\lambda d_{ij} + (1-\lambda).
		\label{effdist}
	\end{equation*}
Tuning $\lambda$ toward 1 represents networks where the cost of moving along a link is strongly \textbf{spatial} (for example, a road network) while choosing $\lambda$ closer to 0 leads to more \textbf{non-spatial} networks (for example, air transportation where the convenience of traveling a route depends more on the number of hops or legs than on the total spatial distance). To illustrate the effect of $\lambda$, we draw two example networks in Fig.~\ref{samecoords}.
Finally, the travel distance is defined as the mean shortest effective path length between all pairs of nodes in the network. 
Taken together, we seek to build networks that minimize travel distance while remaining under a fixed construction budget, i.e.\ given fixed node positions, links are added according to the constrained optimization problem
%
\begin{equation}\label{optim}
\begin{gathered}
     \mathrm{min} \; \f{1}{\binom{N}{2}} \sum_{s,t \in V}{ \sum_{(u,v) \in \Pi(s,t)}{\deff(u,v)} }\\
     \text{subject to} \sum_{(i,j) \in E}{d_{ij} } \leq \text{Budget},
\end{gathered}
\end{equation}
where $\Pi(s,t)$ is the set of links in the shortest effective path between nodes $s$ and $t$, according to the effective distances $\deff$.
This optimization was solved using simulated annealing (see App.~\ref{app:optnetdeets} for details) with a budget of 10 (as in~\cite{gastner2006shape,gastner2006spatial}) and a size of $N$ = 50 nodes. We focus on such a small number of nodes to better mimic realistic microgrid scales. In this work, to average results, 100 individual network realizations were constructed for each $\lambda$.

\begin{figure}
    \includegraphics[width=8.5cm,trim=10 0 35 12, clip=true]{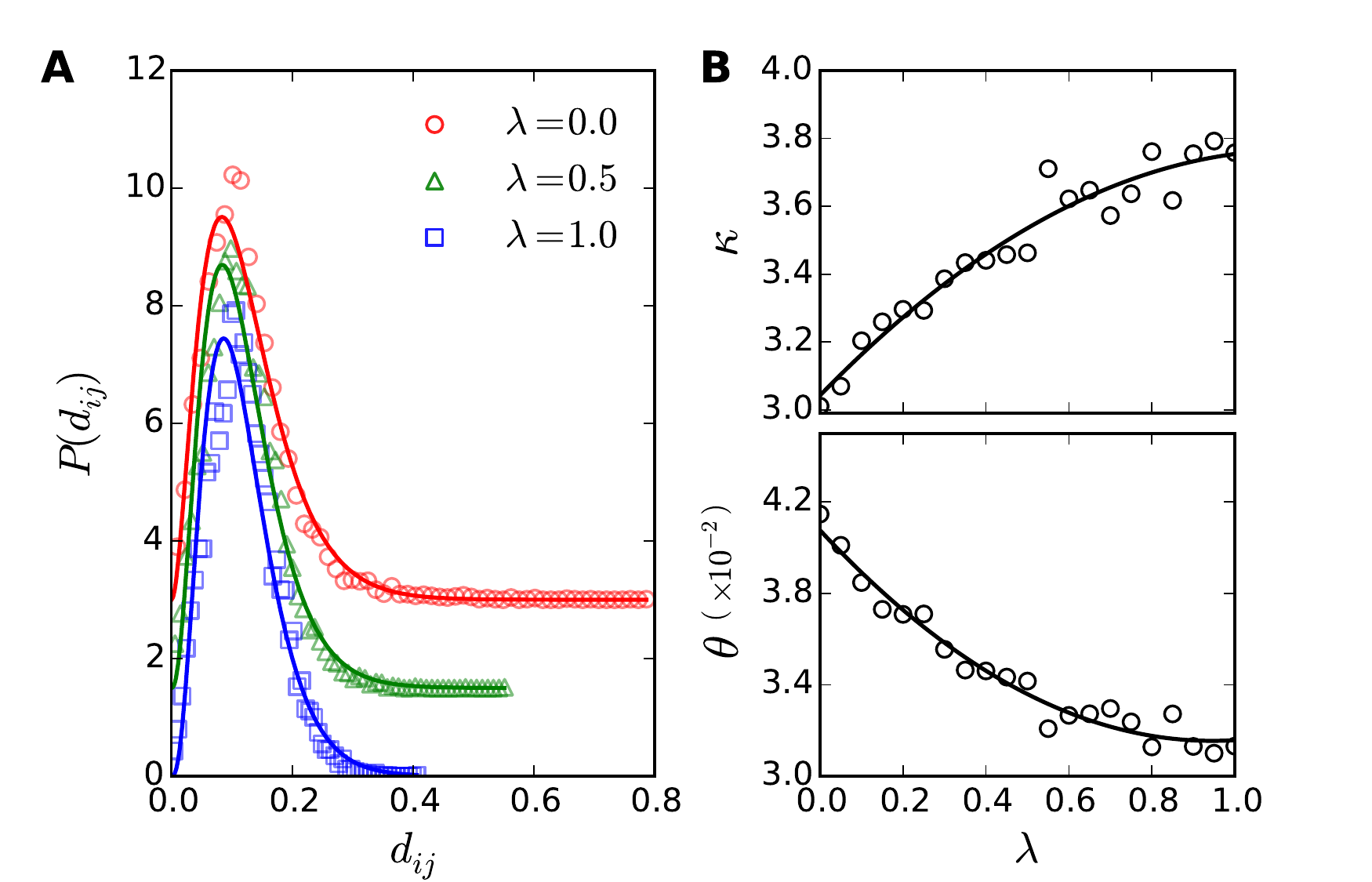}
    \caption{(Color online) The distributions of Euclidean link lengths $d_{ij}$ between nodes $i$ and $j$ are well explained for all $\lambda$ by gamma distributions, i.e.\ $P(d_{ij}) \propto d_{ij}^{\kappa-1}e^{-d_{ij}/\theta}.$ (\textbf{A}) Maximum likelihood estimates of $P(d_{ij})$ for multiple $\lambda$. Two distributions are \textbf{shifted vertically} for clarity. (\textbf{B}) The gamma parameters $\kappa,\theta$ as a function of $\lambda$. Quadratic fits provide a guide for the eye.
    \label{gammadistru}}
\end{figure}

An important quantity to understand in these networks is the distribution of Euclidean link lengths. If edges were placed randomly between pairs of nodes, the lengths would follow the square line picking distribution with mean distance $\avg{d} \approx 0.52141$~\cite{weisstein2005square}. Instead, the optimized network construction makes long links costly and
we observe (Fig.~\ref{gammadistru}) that the probability distribution $P(d)$ of Euclidean link length $d$ after optimization is well explained by a gamma distribution, meaning
the probability that a randomly chosen edge has length $d$ is
\begin{equation}
    P(d) = \f{1}{\Gamma(\kappa)\theta^{\kappa}}d^{\kappa-1}e^{-d/\theta}\label{origdist},
\end{equation}
with shape and scale parameters $\kappa > 0$ and $\theta > 0$, respectively. A gamma distribution is plausible for the distribution of link lengths because it consists of two terms, a power law and an exponential cutoff. This product contains the antagonism between the minimization and the constraint in Eq.~\eqref{optim}: Since longer links are generally desirable for reducing the travel distance, a power law term with positive exponent is reasonable, while the exponential cutoff captures the need to keep links short to satisfy the construction budget and the fact that these nodes are bounded by the unit square. See Fig.~\ref{gammadistru}.

The network parameters were chosen under conditions that were general enough to apply to any small network, for instance a microgrid in a small residential neighborhood. The choices of 50 nodes and a budget of 10 were also made in line with previous studies of this network model to balance small network size with a budget that shows the competition between travel distance minimization and construction cost constraint~\cite{gastner2006shape,gastner2006spatial}. 



\section{Robustness of physical infrastructure}
\label{sec:robustPhysInf}

\begin{figure}
    {\includegraphics[width=8cm,trim=10 10 8 8,clip=true]{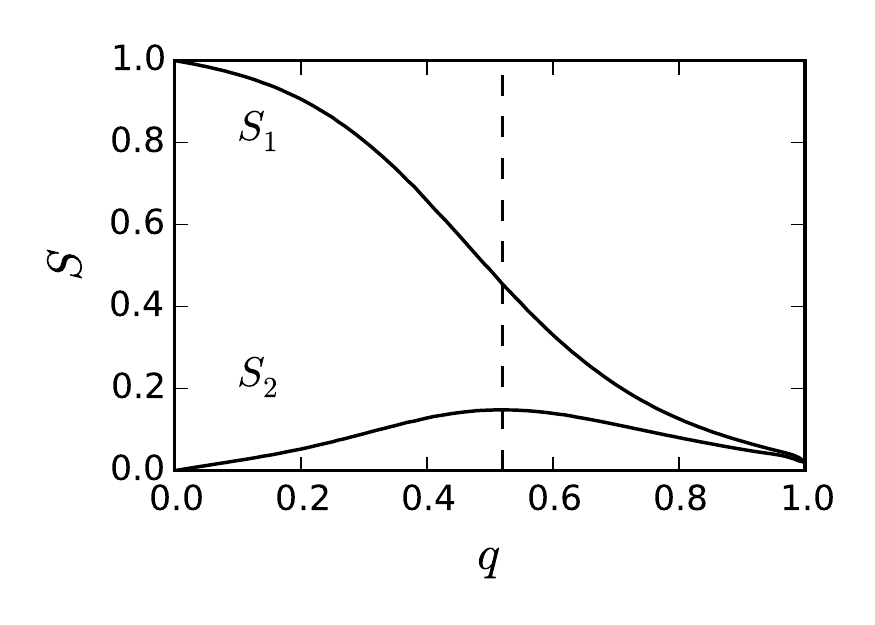}}
    \caption{ The fractions of nodes in the first and second largest components, $S_1$ and $S_2$, respectively, as a function of link deletion probability $q$. In finite systems the percolation threshold $\qc$ can be estimated from the maximum of $S_2$ (dashed line). This example used optimized networks with $\lambda = 1/2$.
    \label{s1s2}}
\end{figure}

Percolation theory on networks studies how networks fall apart as they are sampled. For example, in traditional bond percolation each link in the network is independently retained with probability $p$ (equivalently, each link is deleted with probability $q=1-p$). This process represents random errors in the network. The percolation threshold $\qc$ is the value of $q$ where the giant component, the connected component containing the majority of nodes, first appears. Infinite systems exhibit a  phase transition at $\qc$, which becomes a critical point~\cite{stauffer1991introduction}.
In this work we focus on small \textbf{micronetworks}, a regime under-explored in percolation theory and far from the thermodynamic limit invoked by most analyses. In our finite graphs, we estimate $\qc$ as the value of $q$ that corresponds to the largest $S_2$, where $S_n$ is the fraction of nodes in the $n^{\text{th}}$ largest connected component (Fig.~\ref{s1s2}). In finite systems the second largest component peaks at the percolation threshold; for $q > \qc$ the network is highly disconnected and all components are small, while for $q < \qc$ a giant component almost surely encompasses most nodes and $S_2$ is forced to be small. 
Note that it is also common to measure the average component size excluding the giant component~\cite{hoshen1976percolation,stauffer1991introduction}.

\begin{figure}
    {\includegraphics[width=8cm]{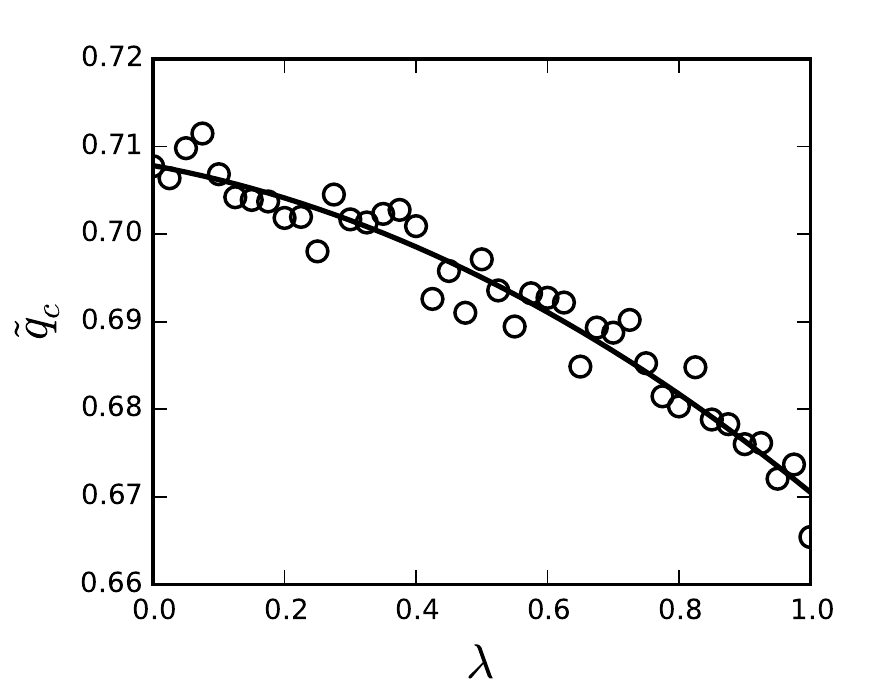}}
    \caption{For an infinite, uncorrelated network, percolation occurs at the sampling probability for which  $\avg{k^2} / \avg{k} = 2$~\cite{cohen2000resilience}. 
We computed this predicted critical point $\tilde{\qc}$ for each $\lambda$ finding $\tilde{\qc}$ between 0.66--0.72. The quadratic fit provides a guide for the eye.
    \label{theoryqcpred}}
\end{figure}

For the case of uniformly random link removals (bond percolation) it was shown that the critical point occurs when $q$ is such that $\langle k^2\rangle / \avg{k} = 2$ ~\cite{cohen2000resilience,molloy1995critical}, where $\avg{k}$ and $\langle k^2 \rangle$ are the first and second moments of the percolated graph's degree distribution, respectively. We denote this theoretical threshold as $\tilde{\qc}$ to distinguish this value from the $\qc$ estimated via $S_2$. Computing this theoretical prediction for the optimized networks (Sec.~\ref{sec:infrastructureModel}) we found $\tilde{\qc}$ between 0.66 and 0.71  for the full range of $\lambda$ (Fig.~\ref{theoryqcpred}). 
It is important to note that the derivation of this condition for $\tilde{\qc}$ makes two related assumptions that are a poor fit for these optimized spatial networks. First, the theoretical model studies networks whose nodes are connected at random. This assumption does not hold for the constrained optimization (Eq.~\eqref{optim}) we study. Second, this calculation neglects loops by 
assuming the network is very large and at least locally treelike. For the small, optimized networks we build this is certainly not the case.
These predictions for the critical point $\tilde{\qc}$ do provide a useful baseline to compare to the empirical estimates of $\qc$ via $S_2$.


\section{Modeling infrastructure robustness}
\label{sec:modelInfRob}

The work by Gastner and Newman~\cite{gastner2006shape} showed the importance of incorporating spatial distances into the construction of an infrastructure network model. With physical infrastructure we argue that it is important to also consider spatial distances when estimating how robust a network is to random failures. For example, consider
%
%
%
a series of power lines built in a rural area where trees are scattered at random.
In a storm trees may fall and damage these lines, and one would expect, all else being equal, that one line built twice as long as another would have twice the chance of a tree falling on it and thus failing. 

\begin{figure*}
    \includegraphics[width=0.8\textwidth,trim=0 15 0 0,clip=true]{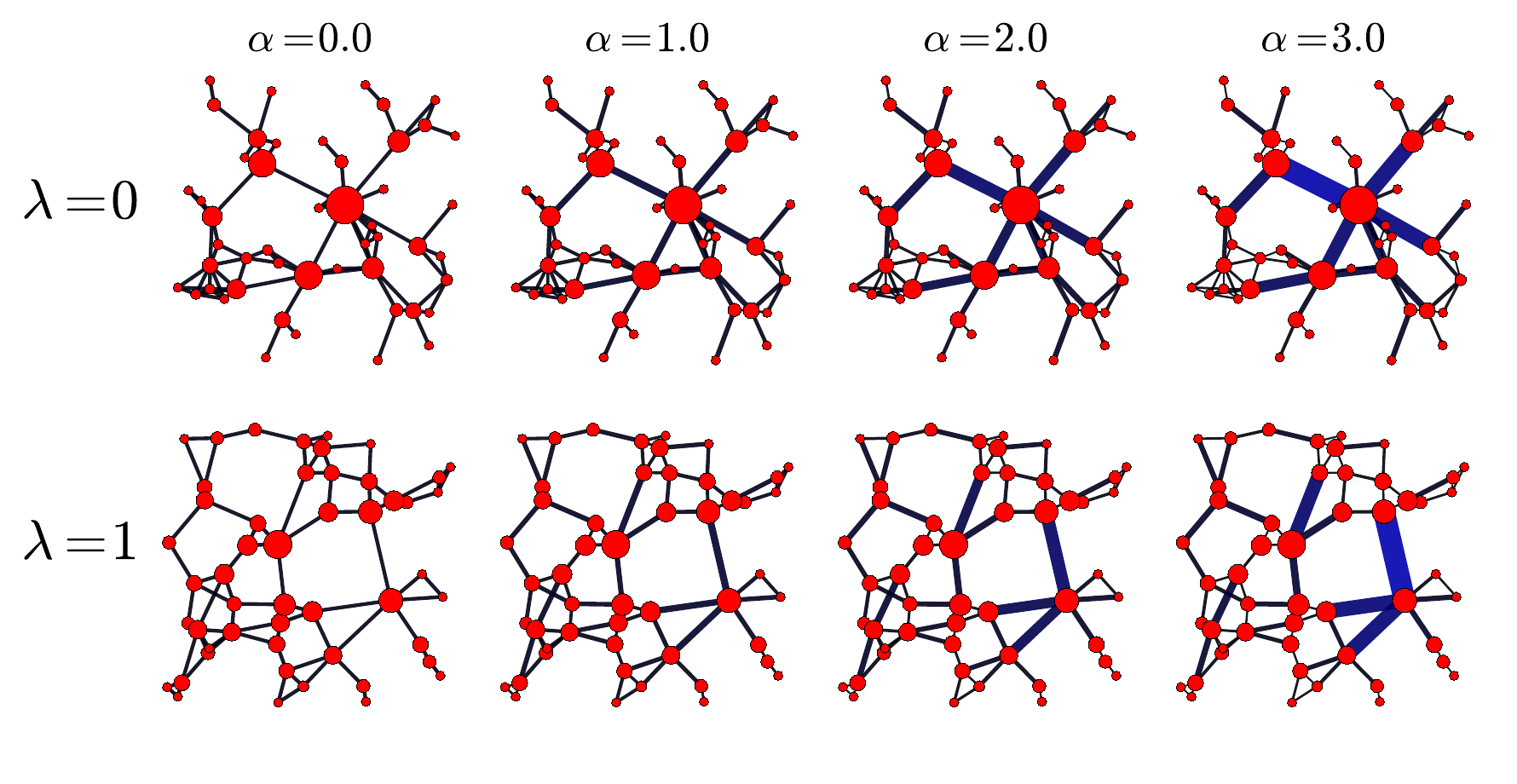}
        \caption{(Color online) A non-spatial ($\lambda=0$) and spatial ($\lambda=1$) network with multiple values of $\alpha$ showing how to tune the role spatial distance plays in percolation.
    Here the width and color of a given edge $(i,j)$ are proportional to failure probability $Q_{ij} \sim d_{ij}^{\alpha}$ \eqref{eqpercprob} and node size corresponds to the number of effective shortest paths through nodes, with the same scales used across all network diagrams.
        %
        Increasing $\alpha$ leads to failure probability becoming more concentrated on the links connected to a small number of hubs, with the effected hubs being more central (in terms of shortest paths) for the non-spatial network ($\lambda = 0$) than for the more geometric network.
        %
        \label{sumryedgeatck}}
\end{figure*}

Motivated by this example, an intuitive model for how links fail would require an increasing chance of failure with length. The simplest model supposes that the failure of a link is directly proportional to length, i.e., that each unit length is equally likely to fail. With this in mind we now introduce the following generalization of bond percolation:
Each link $(i,j)$ independently fails with probability $\min\left(1,Q_{ij}\right)$, where
\begin{align}
    Q_{ij} &= q M\f{d^{\alpha}_{ij}}{\sum_{(i,j)\in E}{d^{\alpha}_{ij}} } = q\f{d^{\alpha}_{ij}}{\avg{d^{\alpha}}},
    \label{eqpercprob}
\end{align}
 $q \in [0,1]$ is a tunable parameter that determines how many edges from $0$ to $|E| = M$ will fail on average, and the parameter $\alpha$ controls how distance affects failure probability. We naturally recover traditional bond percolation ($Q_{ij} = q$) when $\alpha=0$ and $\alpha=1$ corresponds to the case of constant probability per unit length. See Fig.~\ref{sumryedgeatck} for example networks illustrating how $Q_{ij}$ depends on $d_{ij}$ and $\alpha$.

Given the gamma distribution of link lengths, the distribution of $d^\alpha$ is (when $\alpha > 0$)
\begin{equation}
    P(d^{\alpha}) = \f{1}{\alpha\Gamma(\kappa)\theta^{\kappa}} d^{\kappa/\alpha-1} \exp\l(-\f{d^{1/\alpha}}{\theta} \r)
    \label{generaldis}
\end{equation}
 with mean
\begin{equation}
    \avg{d^{\alpha}} =  \f{1}{\alpha\Gamma(\kappa)\theta^{\kappa}} \int_{0}^{\infty} {y^{\kappa/\alpha}} e^{-y^{\f{1}{\alpha}}/\theta } \; \mathrm{dy}
    = \theta^{\alpha} \f{\Gamma(\kappa+\alpha)}{\Gamma(\kappa)}.\label{meangen}
\end{equation}
When $\alpha=1$, the above distribution \eqref{generaldis} will reduce to the original distribution $P(d)$, Eq.~\eqref{origdist}.
%
%
%
%

%

With the above failure model and the distribution $P(d)$, we may express the probability $P(Q_{ij})$ that a randomly chosen edge $(i,j)$ has failure probability $Q_{ij}$ as 
\begin{equation}
    P(Q_{ij}) = \f{1}{\alpha \Gamma(\kappa)\theta^{\kappa}}\l(\f{\avg{d^{\alpha}}}{q}\r)^{\kappa/\alpha} Q_{ij}^{\frac{\kappa}{\alpha}-1} \exp\left[{ -\f{1} {\theta }\l( \f{\avg{d^{\alpha}}}{q}Q_{ij} \r)^{\f{1}{\alpha}}} \right].
\end{equation}
This distribution has mean $\avg{Q_{ij}} = q$. (However, the true mean failure probability is $\avg{\min(1,Q)} \leq \avg{Q}$ which leads to a small correction, easily computed, as $q$ gets closer to 1.)
Note that, while the mean does not rely on the distances of edges, $\alpha$ (and $\avg{d^\alpha}$) do play a role in higher moments. For example, the variance of $Q$ is $\sigma^2(Q) = q^2\left(\mathrm{B}(\alpha,\kappa)/\mathrm{B}(\alpha,\alpha+\kappa)-1 \right)$, where $\mathrm{B}(x,y)$ is the Beta function.

\begin{figure*}
    {\includegraphics[trim=10 8 6 5,clip=true,width=0.8\textwidth]{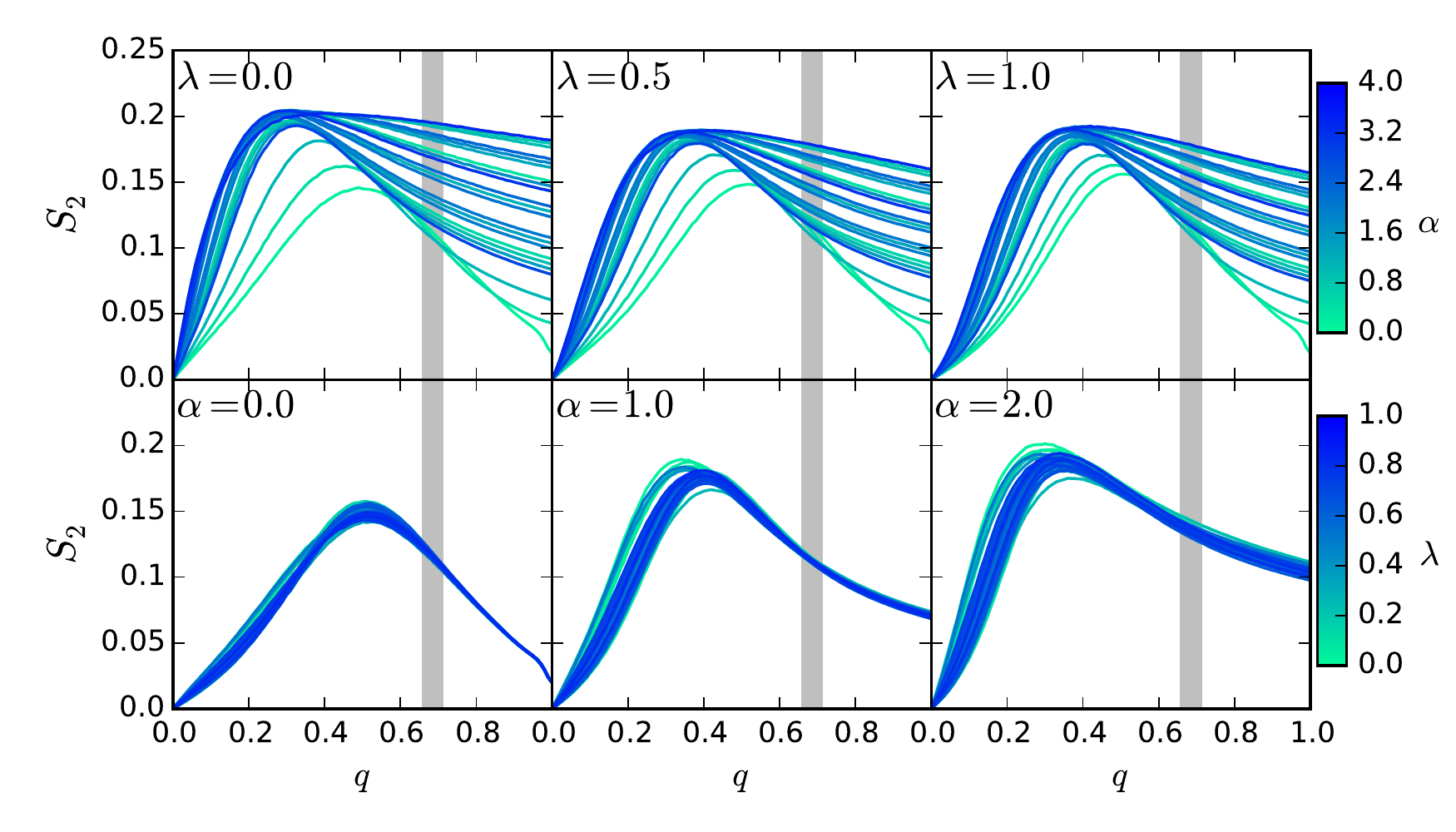}}
    \caption{(Color online) In each panel, $S_{2}$ (fraction of nodes in the 2nd largest component) curves are shown with the range of the theoretical threshold $\tilde{\qc}$ shown in gray.
    Higher values of $\alpha$ make failures depend more strongly on distance, while changing $\lambda$ adjusts a network's form from non-spatial (small $\lambda$) to spatial (large $\lambda$).
    (\textbf{Top row}) Regardless of $\lambda$, larger values of $\alpha$ tend to shift the peak of $S_2$ towards lower $q$, leading to less robust networks. 
    (\textbf{Bottom row})   
    Different values of $\lambda$ for a given $\alpha$ lead to shifted $S_2$ profiles, but the shift is less prominent.
    Regardless of the parameters, spatial networks are more fragile than predicted from theory~\cite{cohen2000resilience,cohen2001breakdown}. While both parameters influence the robustness of these spatial networks, $\alpha$ plays a stronger role than $\lambda$. Note that with our definition of $Q$ \eqref{eqpercprob}, $S_2$ may remain finite as $q \to 1$.\label{s2grid}}
\end{figure*}

To study this robustness model we percolate the infrastructure networks by stochastically removing links $(i,j)$ with probabilities $Q_{ij}$ (Eq.~\eqref{eqpercprob}) for $0<q<1$ and $0 < \alpha < 4$.
In Fig.~\ref{s2grid} we plot $S_2$ vs.\ $q$ for various combinations of $\alpha$ and $\lambda$. 
Importantly, in all cases $\qc < \tilde{\qc}$, indicating that these networks are less robust than predicted. When comparing the effects of each parameter, $\alpha$ has a much greater effect in reducing $\qc$ than $\lambda$; sampling by distance plays a much greater role in determining robustness than how the network is constructed.


\begin{figure}
    \includegraphics[trim=10 0 0 0,clip=true]{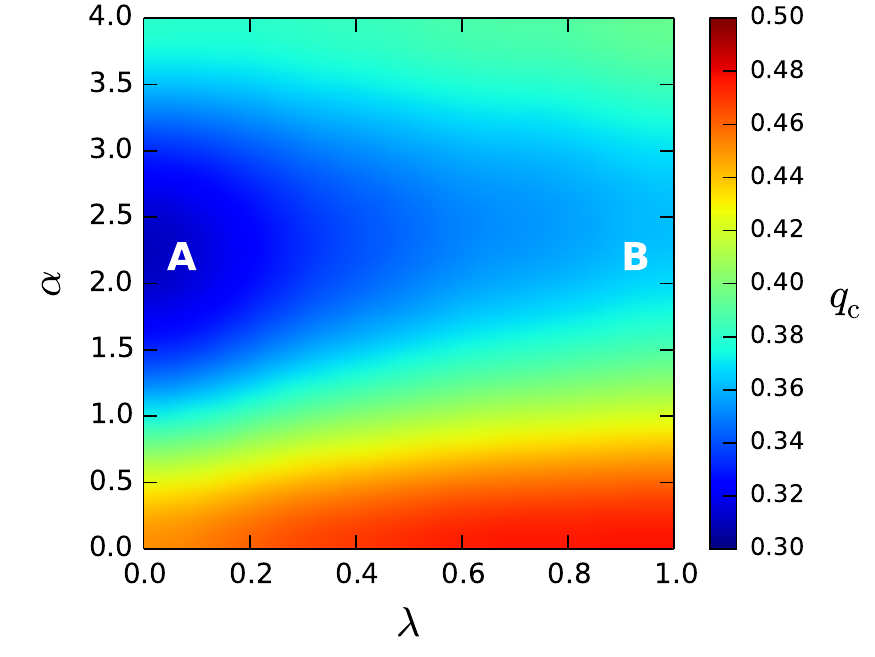}
    \caption{(Color online) Critical failure probability $\qc$ as a function of $\alpha$ and $\lambda$. Overall, values of $\alpha > 0$ always correspond to lower robustness than when $\alpha = 0$ and in particular, the percolation threshold, $\qc$, is lowest near \textbf{A} ($(\lambda,\alpha) \approx (0,2)$), while networks are generally most robust when $\alpha \leq 0.5$. 
    The exponent $\alpha$ lowers $\qc$ even in geometric networks (high $\lambda$) where spatial distance plays a stronger role in the network topology (region \textbf{B}).
    This matrix was smoothed with a $\sigma=$ 5-pixel gaussian convolution for clarity.\label{lvlplot} } 
\end{figure}

The curves in Fig.~\ref{s2grid} show $S_2$ for the entire range of $q$; to study $\qc$ requires examining the peaks of these curves. Figure~\ref{lvlplot} systematically summarizes $\qc$ as a function of $\lambda$ and $\alpha$. Over all parameters, $\qc$ ranges from approximately 0.30 to 0.50. Globally, the most vulnerable region is at \textbf{A} ($(\lambda,\alpha) \approx (0,2)$); these non-spatial networks with strong, super-linear ($\alpha > 1$) failure dependence on distance occupy the most vulnerable region of (Fig.~\ref{lvlplot}) since their construction (low distance dependence) is in direct opposition to how links fail (high distance dependence). Even when networks are built with the goal of minimizing physical distances along links
(high $\lambda$), the exponent $\alpha$ still lowers $\qc$ compared with the theoretical prediction (highlighted at region \textbf{B}). 
Almost any introduction of spatial dependence on link failure (compare $\alpha > 0$ with $\alpha = 0$) leads to less robust networks.






\begin{figure}
    \includegraphics[width=7.7cm,trim=10 10 5 8,clip=true]{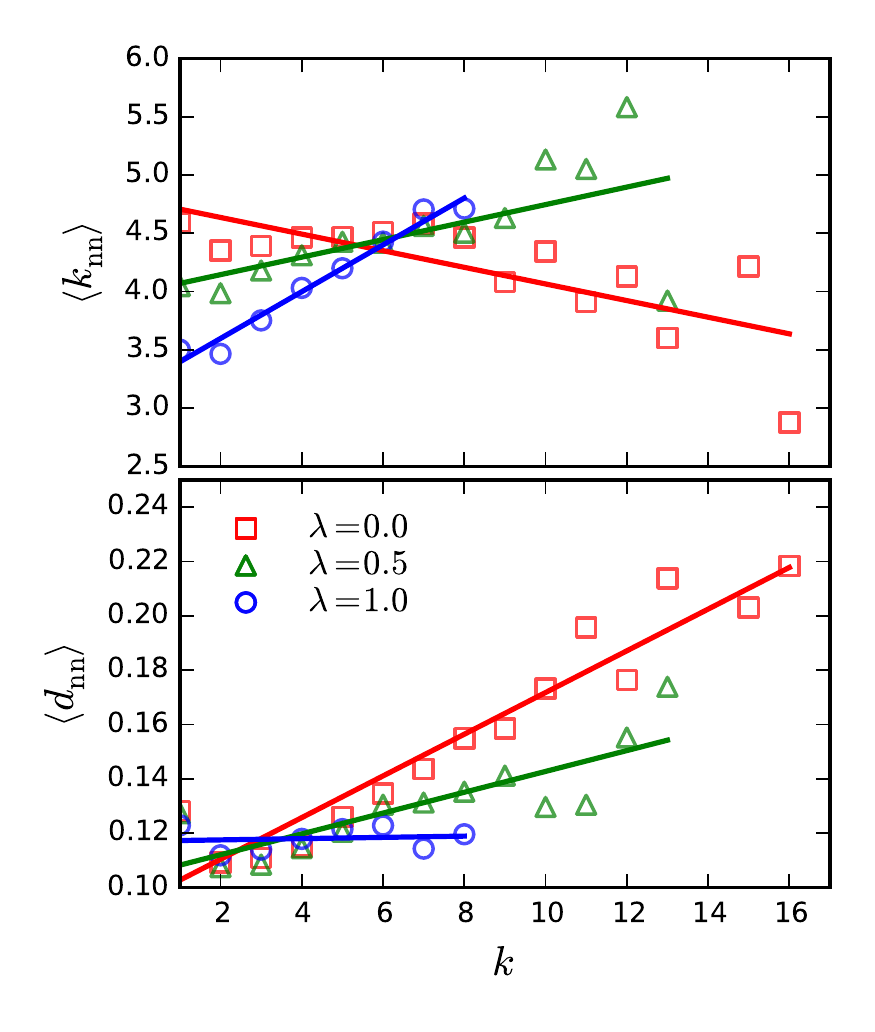}
    \caption{(Color online)
    Degree and distance correlations in optimized spatial networks.
    Here $\knn$ is the mean degree of nearest neighbors and $\dnn$ is the mean distance of nearest neighbors.
    We observe that $\knn$ shows a negative trend with degree $k$ for $\lambda=0$, and positive trend for $\lambda = 0.5$ and $1.0$. On the other hand, $\dnn$ shows an increasing trend with $k$ for decreasing $\lambda$. These optimized networks are not randomly constructed; they possess correlations in either network or spatial structure (or both) for all $\lambda$.  The above metrics indicate that non-spatial networks form hubs whose longer links are likely to fail with higher probability and cause more damage to the network. Alternatively, more spatially-dependent networks (higher $\lambda$) have $\dnn$ that depends less on $k$, indicating that link failures are spread somewhat more uniformly across high- and low-degree nodes. \label{knndnn}}
\end{figure}

Finally, to better understand why these infrastructure networks are less robust than the theoretical prediction~\cite{cohen2000resilience,cohen2001breakdown}, 
we studied correlations in network structure by computing the mean degree of nearest neighbors $\knn =  \sum_{k^\prime}{P(k^\prime \given k)}$~\cite{PhysRevLett.87.258701} and the mean distance to nearest neighbors
$\dnn = \int{d^\prime P(d^\prime \given k)} \; dd^\prime$, both as functions of node
degree $k$. Here $P(k^\prime|k)$ is the 
conditional probability that a node of degree $k$ has a 
neighbor of degree $k^\prime$ and $P(d^\prime|k)$ is the conditional 
probability that a node of degree $k$ has a link of length between $d^\prime$ and $d^\prime + d d^\prime$.
See Fig.~\ref{knndnn}. 
Due to the optimization (Eq.~\ref{optim}), both $\dnn$ and $\knn$ 
indicate non-random structure, since they depend on $k$. Even for the case 
$\lambda = 1$, which shows no relationship between
$\dnn$ and $k$, there is a positive trend for $\knn$. 
 Therefore, the optimized networks always possess correlated topologies. 

Taken together, Fig.~\ref{knndnn} shows that, beyond finite-size effects, $\tilde{\qc}$ overestimates $\qc$  because (i) these networks are non-random and (ii)
higher degree nodes tend to have longer links leading to hubs that suffer more damage when $\alpha > 0$. Since hubs play an outsized role in holding the network together, the positive correlation between $d$ and $k$ causes spatial networks to more easily fall apart, lowering their robustness.



\section{Discussion}
\label{sec:discuss}


A potential application of this model is to designing microgrids. The microgrid concept, most commonly implemented in military settings, has gained wider popularity with the advent of the smart grid. Building a microgrid that is robust to failures while constrained by a budget is important for the widespread adoption of microgrids. Furthermore, the model also brings to light the need to keep in mind that the construction of convenient, long power lines may not be an optimal choice when accounting for the system's robustness. This may reinforce distributed generation across many buildings, as opposed to the power grid (traditional utility) creation of power lines stemming from a centralized cluster of small power plants. A move toward distributed generation and the decommissioning of the traditional utility may raise the overall stability of the grid. Existing infrastructure can use methods that reduce the power grid's dependence on distance (effectively lowering $\alpha$), such as using towers to raise long-distance transmission lines above trees.
Distributed generation may be a cost-effective alternative.

Of course, the metrics used here are not all-encompassing for quantifying robustness. Additional measures may be used that go beyond the topological connectivity of networks to network functionality and dynamics, including problem-specific analyses~\cite{hines2009large,cotilla2012comparing}.
One specific example: it is worth understanding how a spatial network's travel distance may change following link failures, even when a giant component remains.
It is also worth further characterizing fluctuations in, e.g., $\qc$ that are due to the small size of these micronetworks.

Additional future work may include considering the unit square to have differential terrain, changing the cost of edge placement over a continuous gradient. Also, applying the existing model to real infrastructure network data, we may measure robustness of critical networks and have better insight on how to design and improve these structures. Furthermore, in a real power grid nodes do not all have equal roles and thus investigating not only spatially-dependent edge failure but variations in node importance may gain more insight into spatial network robustness. 

\section*{Acknowledgments}
We thank M.~T.~Gastner, J.~R.~Williams, N.~A.~Allgaier, P.~Rezaei, P.~D.~Hines, and P.~S.~Dodds for useful discussions.
This research funded and supported by the National Science Foundation's IGERT program (DGE-1144388), Vermont Complex Systems Center, and the Vermont Advanced Computing Core, which is supported by NASA (NNX 06AC88G). 

\appendix

\section{Constructing optimized networks} \label{app:optnetdeets}


Networks are initialized by first placing $N = 50$ nodes uniformly at random inside the unit square. Initially the network is empty. The minimum spanning tree (MST) is inserted between these nodes using Kruskal's algorithm~\cite{kruskal1956shortest} with link weights corresponding to $\deff$, and the construction cost and travel distance are computed. The spanning tree, which may be modified as optimization progresses, ensures the travel distance is finite when optimization begins. 
%
%
%
We find solutions to the constrained optimization problem (Eq.~\eqref{optim}) using simulated annealing (SA). At the beginning of each SA step, an edge is added to the network at random and construction cost and travel distance are recomputed. If the budget constraint is still satisfied with the addition of this edge, the edge is kept using Boltzmann's criterion: the edge is retained if it lowers the travel distance; if it does not lower the travel distance it is retained with probability $e^{-\beta \Delta E}$, where $\Delta E$ is the change in travel distance due to this change in the network, and $\beta$ acts as the inverse temperature. 

If the random edge puts the network over budget, we remove it and do one of two modifications. With probability one half an existing edge is moved by placing it at random in the network where no edge exists. Otherwise, a rewire is chosen. Edges are rewired by first selecting an existing edge at random, next selecting either of the nodes connected by that edge, and finally attaching that end of the edge from the chosen node to a node that is a non-neighbor. In other words, edge $(i,j)$ is removed and edge $(i,k)$ is inserted where $k\neq j$ and $k$ was not previously a neighbor of $i$. 
The move/rewire perturbation is then kept using the same Boltzmann's criterion.

%
%
%

The cooling schedule starts at $\beta_0 =100/(\text{cost of MST})$, and cooled subsequently as $\beta_{t+1} = \beta_t \l(1+3 \times 10^{-5}\r)$.
At each SA step we check if the current network topology is the best seen to that point; the most optimal network found during any of the $3\times 10^5$ total SA steps is taken as our solution.

\end{document}